\documentstyle[twocolumn,aps,psfig,amssymb]{revtex}
\begin{document}
\title{Thermodynamic Properties of Strongly Degenerate Interacting Fermi Systems}
\author{V.S.~Filinov\footnote{Permanent address: Russian Academy of Sciences,
High Energy Density Research Center,
Izhorskaya street 13-19, Moscow 127412, Russia} and M.~Bonitz}
\address{Fachbereich Physik, Universit{\"a}t Rostock\\
Universit{\"a}tsplatz 3, D-18051 Rostock, Germany}
\date{\today}
\maketitle
\begin{abstract}
A numerical approach is presented which allows to
calculate the equilibrium properties of Fermi systems
which are both, {\em strongly coupled and strongly degenerate.}
Based on a novel path integral representation of the
many-particle density operator, quantum and spin effects are included.
As an illustration, results for the pressure and energy of an
electron--proton plasma are presented.
\end{abstract}
\pacs{  }

Coulomb systems continue to attract the interest of researchers in many
fields, including plasmas, astrophysics, solids and nuclear matter, see
Refs.~\cite{boston97,binz96} for an overview. Stimulated by recent impressive
experimental advances \cite{boston97}, theoretical activities follow a large
variety of approaches: exactly solvable models \cite{boston97,binz96},
kinetic theory and quantum statistics, e.g. \cite{green-book,bonitz-book},
computer simulations, e.g. \cite{zamalin,binder96,berne98} and others.
The most interesting phenomena
such as metallic hydrogen, plasma phase transition, Fermi liquids,
bound states etc., occur in situations where both
Coulomb {\em and} quantum effects are relevant. However, here,
all present theories have practical difficulties. Quantum kinetic
methods easily handle quantum and spin effects, but they become extremely
complicated when correlation effects can not be treated perturbatively.
On the other hand, Monte Carlo (MC) and molecular dynamics (MD) simulations
allow for an
efficient treatment of strong coupling phenomena, but still have difficulties
with incorporating quantum and spin statistics effect.

In particular, there has been remarkable progress in applying path integral
quantum Monte Carlo (PIMC) techniques to Bose \cite{ceperley95rmp} and Fermi
systems, see e.g.
\cite{zamalin,binder96,berne98,pierleoni-etal.94} and
Ref.~\cite{ceperley95} for an
overview. However, there remains one major obstacle preventing efficient
modelling of Fermi systems -- the so-called sign problem, i.e. the appearance
of differences of large numbers in the expressions for thermodynamic quantities.
It is related to the antisymmetrization of the fermionic N-particle density
matrix which results in a sum of $N!/2$ positive and negative terms in the
thermodynamic functions. This led to a
number of additional approximations, such as the fixed node and
restricted path integral concepts. Despite
impressive recent simulation results, see \cite{pierleoni-etal.94,ceperley95}
and references therein, to overcome these assumptions remains one of the
challenges from the fundamental (first principal character) as well as practical
point of view.

In this work we demonstrate that these approximations can in fact be avoided.
We derive a different path integral representation for the density matrix $\rho$
by introducing a modified set of integration variables. This leads to two major
advantages for the thermodynamic functions: i) they do not contain
an explict sum over permutations (with alternating sign) and ii) part of the
volume and temperature dependencies of $\rho$ can be eliminated. As an example,
we present explicit formulas for the equation of state and energy of a
two-cmponent plasmas, cf. Eqs.~(\ref{eos}) and (\ref{energy}).
The numerical efficiency of our approach is demonstrated
with results for the pressure and energy of a
high-temperature/high-density electron-proton plasma over a wide range of
degeneracy parameters. Our approach has important consequencies for
simulations of Fermi systems, such as dense quantum
plasmas, few-fermion systems in traps or quantum confined semiconductor
structures \cite{afilinov-etal.99prl} and is the basis for developing
path integral MD simulations for Coulomb systems \cite{filinov-kbt99}.

As is well known the thermodynamic properties of a quantum system of $N$
particles are fully
determined by the partition function Z and, consequently, by the density matrix
\cite{feynman-hibbs}
\begin{eqnarray}
Z = \int\limits_{V} dq \,\rho(q,0;q,\beta), \quad
\rho(q,\beta;q',\beta') \equiv \langle q \beta|{\hat \rho}|\beta' q'\rangle ,
\label{z-rho}
\end{eqnarray}
where ${\hat \rho}=\exp\{-(\beta-\beta') {\hat H}\}$,
$\beta=1/k_BT$, and
$q$ comprises the coordinates of all particles,
$q \equiv \{{\bf q}_1, {\bf q}_2,..., {\bf q}_N\}$. With an analytical
expression for the density matrix given, one can use e.g. Monte Carlo methods
\cite{zamalin,binder96,berne98,ceperley95} to evaluate the partition function and
thermodynamic quantities. However, for a quantum system, $\rho$ is, in
general, not known but can be constructed from its known high-temperature
limit by means of a decomposition into $n+1$ factors, each of which
corresponds to the density matrix at an $n+1$ times higher temperature
\cite{feynman-hibbs},
\begin{eqnarray}
\rho(q,0;q',\beta) = \frac{1}{N!} \sum_{P} (\pm
1)^{\kappa_P} \int\limits_{V} dq^{(1)} \dots dq^{(n)} \,
\nonumber \\
\times \rho(q,0;q^{(1)},\tau^{(1)}) \, \dots \,
\rho(q^{(n)},\tau^{(n)};{\hat P}\ q',\beta),
\label{rho-sym}
\end{eqnarray}
where $\tau^{(i+1)}-\tau^{(i)}=\Delta \beta = \beta/(n+1)$.
Eq.~(\ref{rho-sym}) lets one view the N-particle state as a loop consisting of
n+1 vertices (``beads'') located at intermediate coordinates
$q^{(i)}, \tau^{(i)}, i=1 \dots n$.
For quantum systems of bosons (fermions), furthermore, in Eq. (\ref{rho-sym}),
the spin statistics theorem has been taken into account which requires to
perform an (anti-)symmetrization of the density matrix giving rise to the sum over all
$N$-particle permutations $P$ with parity $\kappa_P$, and ${\hat P}$
denotes the permutation operator. Obviously, for fermions
(minus sign in the parantheses), this sum contains $N!/2$ positive and $N!/2$
negative terms. (To simplify the notation, we have not written the spin
variables explicitly, their inclusion is straightforward).

The unknown density matrix (\ref{rho-sym}) is efficiently computed by
approximating the $\rho$'s on the r.h.s. of Eq.~(\ref{rho-sym}) by the
high-temperature density matrix $\rho^{hT}$. For potentials which are
bounded from below, the simplest choice is, e.g. \cite{zamalin,binder96},
\begin{eqnarray}
\rho^{hT}[q^{(i)},\tau^{(i)};q^{(i+1)},\tau^{(i+1)}] &\approx&
\rho_0[q^{(i)},\tau^{(i)};q^{(i+1)},\tau^{(i+1)}] \,
\nonumber\\
&\times& e^{-[\tau^{(i)}-\tau^{(i+1)}]U(q^{(i)})},
\nonumber\\
\rho_0[q^{(i)},\tau^{(i)};q^{(i+1)},\tau^{(i+1)}]&=& \frac{1}{\lambda_{\Delta}^{3N}}\,
\prod_{p=1}^{N}
e^{- \frac{\pi}{\lambda_{\Delta}^{2}} \left|q_p^{(i)}-q_p^{(i+1)}\right|^2},
\nonumber\\
&&
\label{rho-ht}
\end{eqnarray}
where $\rho_0$ is the free-particle density matrix which is a product of
single-particle density matrices, $U$ is the potential
energy, and $\lambda_{\Delta}$ is the DeBroglie wave length corresponding to
 $k_B(n+1)T=1/\Delta \beta$,
$\lambda_{\Delta}^{2} \equiv 2\pi\hbar^2 \Delta \beta/m$. It is well known
that for $n\rightarrow \infty$, $\rho^{hT}$ converges to the exact
density matrix  $\rho$. Notice that the factorization of $\rho^{hT}$ into
a kinetic and potential energy term, Eq.~(\ref{rho-ht}), is an approximation
also. The error made thereby is of the order of the variation of $U$ on the
spatial scale $\lambda_{\Delta}$ and thus vanishes with $n\rightarrow \infty$.
This holds also for a repulsive Coulomb potential, whereas the
attractive electron-ion interaction has to be represented by a bounded from
below effective pair potential
\cite{green-book,zamalin,ceperley95,pierleoni-etal.94}.

After recalling the general concept of PIMC, we now explain our
scheme. To simplify the presentation, we choose a binary mixture
of quantum electrons and classical protons, $N_e=N_i=N$. The first crucial
step is to transform the intermediate electron coordinates. To this end, we
rewrite the partition function (\ref{z-rho}), now explicitly including the
arguments of the ions,
\begin{eqnarray}
Z(N,V,\beta) &=&
\frac{Q(N_e,N_i,\beta)}{N_e!N_i! \,\lambda_i^{3N_i}\lambda_{\Delta}^{3N_e}},
\nonumber\\
\mbox{with} \qquad
Q(N_e,N_i,\beta) &=& \int\limits_V dq \,dr \,d\xi
\,\rho(q,[r],\beta). \label{q-def}
\end{eqnarray}
Here, the notation $q$ is retained for the ions. $[r]$ summarizes the
electron coordinates with $r$ denoting the coordinates at the beginning
of the fermionic loop and $\xi^{(1)}, \dots \xi^{(n)}$ the dimensionless
distances between neighboring vertices on the loop. Thus, explicitly,
$[r]\equiv [r; r+\lambda_{\Delta} \xi^{(1)};
r+\lambda_{\Delta}(\xi^{(1)}+\xi^{(2)}); \dots]$,
and $q, r, \xi^{(i)}$ each are $3N$-dimensional vectors. For the density
matrix in Eq.~(\ref{q-def}) we have,
using (\ref{rho-ht}), \cite{zamalin,filinov76}
\begin{eqnarray}
\rho(q,[r],\beta) &=& \sum\limits_{s=0}^N \rho_s(q,[r],\beta)
\label{rho-def}
\\
&=& \sum\limits_{s=0}^N \frac{C^s_N}{2^N}\,
e^{-\beta U(q,[r],\beta)} \prod\limits_{l=1}^n
\prod\limits_{p=1}^N \phi^l_{pp}
{\rm det} \,|\psi^{n,1}_{ab}|_s,
\nonumber\\
\mbox{where} &&
 U(q,[r],\beta) = U^i(q)
\nonumber\\
&+&
\frac{1}{n+1 }\sum\limits_{l=0}^n \left\{
U^e_l([r],\beta) + U^{ei}_l(q,[r],\beta)
\right\},
\label{u-def}
\end{eqnarray}
and $U^i$, $U^e_l$ and $U^{ei}_l$ denote the interaction energy between ions,
electrons at vertex ``$l$'' and electrons (vertex ``$l$'') and ions, respectively.
Furthermore, $\phi^l_{pp}\equiv \exp[-\pi |\xi^{(l)}_p|^2]$
arises from the kinetic energy density matrix $\rho_0$ of the electron with
index $p$. Notice that, in contrast to $\rho_0$ in Eq. (\ref{rho-ht}),
$\phi^l_{pp}$ is independent of volume and
temparature. Furthermore, we underline that, in contrast to
Eq.~(\ref{rho-sym}), Eq.~(\ref{rho-def}) does not contain an explicit
sum over the permutations and thus no sum of terms with alternating
sign. Instead, the whole exchange problem is
contained in a single exchange matrix given by
\begin{eqnarray}
||\psi^{n,1}_{ab}||_s\equiv ||e^{-\frac{\pi}{\lambda_{\Delta}^2}
\left|(r_a-r_b)+ y_a^n\right|^2}||_s,
\label{psi}
\end{eqnarray}
where $y_a^n=\lambda_{\Delta}\sum_{k=1}^{n}\xi^{(k)}_a$.
Besides being a function of distance, this matrix contains all spin
depencies of the N-electron subsystem.
The calculation of $det|\psi_{ab}|_s$ which is crucial for
evaluating the density matrix (\ref{rho-def}) is
greatly simplified by its block structure (it contains two zero
sub-matrices) which is explained in Fig. 1. As a result of the spin summation,
the matrix carriers a subscript $s$ denoting the number of electrons having
the same spin projection.

As a first example of applying our results (\ref{q-def}), (\ref{rho-def})
to thermodynamic properties,
we provide the result for the equation of state,
$\beta p = \partial {\rm ln} Q / \partial V =
[\alpha/3V \partial {\rm ln} Q
/ \partial \alpha]_{\alpha=1}$,
\begin{eqnarray}
\frac{\beta p V}{2N} = 1 + \frac{1}{6NQ}\sum_{s=0}^N
\int dq \, dr \, d\xi \,\rho_s(q,[r],\beta) \,\times \qquad
\nonumber\\
\Bigg\{\sum_{p<t}^{N_i} \frac{\beta e^2}{|q_{pt}|} +
\sum_{p<t}^{N_e} \frac{\beta e^2}{(n+1)|r_{pt}|}
-  \sum_{p=1}^{N_i}\sum_{t=1}^{N_e} \frac{|x_{pt}|}{n+1}
\frac{\partial \beta \Phi^{ie}}{\partial |x_{pt}|}+
\nonumber\\
\sum_{l=1}^{n}\left[\sum_{p<t}^{N_e}
\frac{\beta e^2 \langle r^l_{pt}|r_{pt}\rangle }{(n+1)|r^l_{pt}|^3}
- \sum_{p=1}^{N_i}\sum_{t=1}^{N_e}
\frac{\langle x^l_{pt}|x_{pt}\rangle}{(n+1)|x^l_{pt}|}
\frac{\partial \beta \Phi^{ie}}{\partial |x^l_{pt}|}
\right]
\nonumber\\
\,+\,\frac{\alpha}{{\rm det} |\psi^{n,1}_{ab}|_s}
\frac{\partial{\rm det} | \psi^{n,1}_{ab} |_s}{\partial \alpha}
\Bigg\}.
\label{eos}
\end{eqnarray}
Here, $\Phi^{ie}$ is the effective electron-ion pair potential,
$\alpha$ is a scaling parameter for the length, $\alpha = L/L_0$,
$\langle \dots | \dots \rangle$ denotes the scalar product, and
$q_{pt}$, $r_{pt}$ and $x_{pt}$ are differences of two
coordinate vectors:
$q_{pt}\equiv q_p-q_t$,
$r_{pt}\equiv r_{p}-r_{t}$, $x_{pt}\equiv r_p-q_t$, $r^l_{pt}=r_p-r_t+y_p^l$
and $x^l_{pt}\equiv r^l_{p}-q_{t}+y^l_p$.
The structure of Eq.~(\ref{eos}) is obvious: we have separated the classical
ideal gas part (first term). The ideal quantum pressure in excess of the
classical one and the correlation
contributions are contained in the
integral term, where the second line results from the ionic correlations
(first term) and the e-e and e-i interaction at the first vertex (second
and third terms respectively). The third and fourth lines are due to the
further electronic vertices and the explicit
volume dependence of the exchange matrix, respectively.
As a consequence of representation (\ref{rho-def}),
the result (\ref{eos}) for the pressure does not contain a sum over
permutations. In fact, each of the sums in curly brackets in Eq.~(\ref{eos})
is bounded as the number of vertices increases,
$n\rightarrow \infty$, which enables us to evaluate the
pressure without additional approximations for the density
matrix.

Other thermodynamic quantities exhibit the same favorable behavior. As a second
result, we provide the formula for the energy,
$\beta E= 6N/2 -\beta \partial {\rm ln} Q
/ \partial \beta$
\begin{eqnarray}
\beta E = 6Nk_BT + \frac{1}{Q}\sum_{s=0}^N
\int dq \, dr \, d\xi \,\rho_s(q,[r],\beta) \,\times  \qquad
\nonumber\\
\Bigg\{\sum_{p<t}^{N_i} \frac{\beta e^2}{|q_{pt}|} +
\sum_{p<t}^{N_e} \frac{\beta e^2}{(n+1)|r_{pt}|}
+  \sum_{p=1}^{N_i}\sum_{t=1}^{N_e} \frac{\beta\Phi^{ie}(|x_{pt}|)}{n+1}
\nonumber\\
+ \sum_{l=1}^{n}\Bigg[\sum_{p<t}^{N_e}
  \frac{\beta e^2}{(n+1)|r^l_{pt}|}
- \sum_{p<t}^{N_e}
  \frac{\beta e^2 \langle r^l_{pt}|y^l_{pt}\rangle }{2(n+1)|r^l_{pt}|^3}
\nonumber\\
+ \sum_{p=1}^{N_i}\sum_{t=1}^{N_e}
\frac{\beta^2\langle x^l_{pt}|y^l_{p}\rangle}{2(n+1)|x^l_{pt}|}
\frac{\partial \beta \Phi^{ie}(x^l_{pt})}{\partial |x^l_{pt}|}
\nonumber\\
+ \sum_{p=1}^{N_i}\sum_{t=1}^{N_e}
\frac{\beta \Phi^{ie}(x^l_{pt})}{n+1} \Bigg]
\,-\,\frac{\beta}{{\rm det} |\psi^{n,1}_{ab}|_s}
\frac{\partial{\rm det} | \psi^{n,1}_{ab} |_s}{\partial \beta}
\Bigg\},
\label{energy}
\end{eqnarray}
where we denoted $y^l_{pt}\equiv y^l_{p}-y^l_{t}$. The structure of
this result is similar to that for the pressure, Eq. (\ref{eos}). Again,
each sum in the parantheses is bounded for $n\rightarrow \infty$.

Besides their advantageous analytical properties, expressions (\ref{eos})
and (\ref{energy}) are well suited for numerical evaluation using Monte
Carlo techniques.  For a fast generation of a sequence of N-particle
configurations (Markov chain) \cite{binder96,zamalin} it is necessary to
efficiently compute the ratio $R$ of two exchange determinants corresponding
to subsequent configurations of the Markov chain,
$R=det |\psi^{n,1}_{ab}|_{new}/det |\psi^{n,1}_{ab}|_{old}$.
The probability of accepting a MC configuration
is proportional to the absolute value of $R$ while the
sign of the determinants is included in the weight function of each
configuration. Without going into details we only mention that this can be
done using the inverse of the exchange matrix. Moreover, the inverse matrix
is used for an efficient computation of the derivatives of the exchange
determinants in Eqs. (\ref{eos}) and (\ref{energy}) which is
expressed in standard manner as a sum of $N$ determinants involving
derivatives of $\psi^{n,1}_{ab}$, Eq. (\ref{psi}).

To demonstrate our numerical scheme, we consider as an example
a two-component electron-proton plasma. For the potential $\Phi^{ie}$
we chose a simple approximation of the Kelbg potential which is the
high-temperature limit,
of the exact electron-ion interaction, e.g. \cite{green-book,filinov76}.
Thus, by increasing the number of electronic  vertices $n$, in
principle, any desired accuracy can be achieved. To simplify the
computations, we included only the dominant contribution in the
sum over $s$ corresponding to $s=N/2$ electrons having spin up and down,
respectively. (The contribution of the other terms is small and
vanishes in the thermodynamic limit.)

As a first test of our approach,
we consider a mixture of ideal electrons and protons for which the
thermodynamic quantities are known analytically, e.g. \cite{green-book}.
Fig.~2 shows our numerical results for the pressure together with the
theoretical  curve. The agreement, up to values of the degeneracy parameter
$\chi\equiv n\lambda^3$ as large as 5 is remarkable. Even with only
$N=32$ electrons and protons deviations are rather small.
One clearly sees that increasing the number of particles (see inset of
Fig. 2) improves the numerical  results systematically and extends the
range of applicability to larger values of $\chi$.

Let us now turn to the case of interacting electrons and protons. We have
performed a series of calculations in which the classical coupling parameter
$\Gamma=(4\pi n_e/3)^{1/3}e^2/4\pi \epsilon_0 kT$ was kept constant
while the degeneracy was varied. The results for the pressure and
energy are presented in Fig.~3. One can see that for weak coupling and
small degeneracy parameters, $\chi<0.5$, exchange effects are small, and QMC
simulations  without exchange (open circles) are close to our results.
However, with increasing $\chi$ and $\Gamma$, the deviations are growing
rapidly. Even stronger are the discrepancies with analytical theories
which are constructed as perturbation expansions, and thus are limited
to small values of $\chi$ and $\Gamma$. Most strikingly is the decrease
of the energy as a function of $\chi$ predicted by the analytical
models and QMC without exchange, which is in
contrast to our results which show an increase for all values of
$\Gamma$. We mention that for all results shown, the maximum statistical
error is about $5\%$ and can be systematically reduced by increasing the
length of the MC run. Furthermore, the accuracy is affected by the number $n$
of vertices. In our calculations, we considered always temperatures above
one Rydberg, for which we checked that it is sufficient to use $n=6$.

In summary, we presented a new path integral representation (\ref{rho-def})
for the $N-$particle density matrix which does not contain an explicit
summation over permutations. As a result, we were able to compute the pressure
and energy of a strongly correlated plasma of degenerate electrons and
classical protons without additional assumptions for the density matrix.
Our numerical results demonstrate the practical feasibility of our
approach and open the way to a large variety of applications including
dense hydrogen, fermion systems in condensed matter, few-fermion
systems in traps \cite{afilinov-etal.99prl} or semiconductor
nanostructures as well as to MD simulations for fermions \cite{filinov-kbt99}.

We acknowledge support from the Deutsche Forschungsgemeinschaft
(Mercator-Programm) for VSF and stimulating discussions with
B.~Bernu, D.~Ceperley, W. Ebeling, V.E.~Fortov, D.~Kremp and M.~Schlanges.

\begin{figure}
\caption{Schematic of the exchange matrix $||\psi_{ab}||$. It contains two
zero sub-matrices, $s$ denotes the number of electrons with the same
spin projection.}
\end{figure}
\begin{figure}
\caption{Pressure (upper figure) and energy (lower figure) of an {\em ideal} plasma
of degenerate electrons and classical proton. Theory (dashed line) is compared
to PIMC simulations with varying particle number.}
\end{figure}
\begin{figure}
\caption{Pressure (upper figure) and energy (lower figure) of a {\em nonideal} plasma
of degenerate electrons and classical proton. Lines with symbols are PIMC
simulations for different values of $\Gamma$ (see inset in upper figure).
Remaining symbols denote quantum MC simulations without exchange (QMCNE) and
analytical models (RSDWK - quantum statistical model of Rieman et al.),
data from Ref. [16].}
\end{figure}

\end{document}